\documentstyle[sprocl,epsfig,rotate]{article}

\def\Journal#1#2#3#4{{#1} {\bf #2}, #3 (#4)}

\def\ZPC{{\em Z. Phys.} C}
\def\EPJ{{\em Eur. Phys. J.} C}
\def\JPG{{\em J. Phys.} G}

\def\ra{\rightarrow}

\def\be{\begin{equation}}
\def\ee{\end{equation}}
\def\bea{\begin{eqnarray}}
\def\eea{\end{eqnarray}}

\begin{document}

\vspace*{-2.3 cm}
\begin{flushright}
CERN-TH-2000-223  \\
DAMTP-2000-72 
\end{flushright}

\vspace*{0.7 cm}

\title{PHYSICS AT HIGH {\boldmath $Q^2$} AND {\boldmath $p^2_t$} :
  SUMMARY OF DIS 2000}

 \author{M. KUZE}
 \address{Institute of Particle and Nuclear Studies, KEK, Tsukuba
 305-0801, Japan}
 \author{S. LOLA}
 \address{CERN Theory Division, CH-1211, Geneva 23, Switzerland}
 \author{E. PEREZ}
 \address{CEA Saclay, DSM/DAPNIA/Spp, 91191 Gif-sur-Yvette, France}
 \author{B. ALLANACH}
 \address{DAMTP, Wilberforce Rd, Cambridge CB3 0WA, UK}


\maketitle\abstracts{We summarize the experimental and theoretical results
presented in the ``Physics at the Highest $Q^2$ and $p^2_t$" working group
at the DIS 2000 Workshop.
High $Q^2$ and $p^2_t$ processes measured at current and future
colliders allow to improve our knowledge of Standard Model (SM) physics,
by providing 
precise measurements of the SM parameters
and, consequently, consistency checks of the SM.
Moreover, they give information on key quantities for
the calculation of the SM expectations in a yet unexplored domain,
such as the parton densities of the proton or the photon.
In addition to these experimental inputs, higher-order calculations are
also needed to obtain precise expectations for SM processes, which are
a key ingredient for the searches for new phenomena in high
$Q^2$ and $p^2_t$ processes at current and
future experiments.
The experimental and theoretical status of
SM physics at high $Q^2$ and $p^2_t$ is reviewed in the first
part of this summary, with the remaining being dedicated 
to physics beyond the Standard Model.}


\section{Jet Physics at High {\boldmath $p^2_t$} and {\boldmath $Q^2$}}
The production of jets at high-energy colliders is a very good
testing ground of perturbative QCD,
and a tool for determining the parton distribution
functions of the proton and the photon.
Possible deviation from the QCD predictions at the highest energy scales
would be a hint for unrevealed new physics.
A good understanding of exclusive multi-jet final states is also crucial
in many searches for new-particle productions, 
such as  
the production
and decay of $R$-parity-violating squarks.

The status of jet production at HERA was reviewed in
the photoproduction~\cite{REPOND} and DIS~\cite{KEIL} regions.
In photoproduction, dijet cross-sections 
have been
measured up to
$p_{t,jet} \approx 80$~GeV and $m_{jj} \approx 150$~GeV,
and three-jet cross-sections up to $m_{jjj} \approx 200$~GeV.
In the direct-photon regime the measurements are in good agreement
with the NLO QCD predictions, while in the resolved-photon regime
one finds a discrepancy, which could be 
attributed 
to the limited knowledge of the photon structure functions.

\begin{figure}[htb]
 \begin{center}
 \begin{tabular}{p{0.58\textwidth}p{0.37\textwidth}}
     \raisebox{-140pt}{
     \mbox{\epsfxsize=0.6\textwidth
         \epsffile{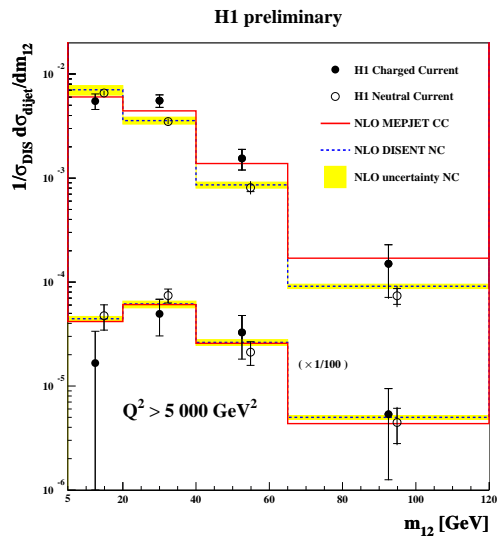}}}
 &
         \caption
         {\small \label{fig:disjet}
            Dijet mass distribution for NC/CC DIS at very high $Q^2$,
            compared to NLO QCD predictions.}
 \end{tabular}
 \end{center}
\end{figure}

In the DIS region, dijet cross-sections can be used to extract
the strong coupling constant $\alpha_s$.
Also, dijet production at very high $Q^2$, $Q^2 \approx 10^4$~GeV$^2$,
has been
studied in both neutral current (NC) and charged current (CC)
processes (Fig.~\ref{fig:disjet}).
Even at these extremely short distances, the NLO QCD predictions
work very well within the statistics of the current data.

At TeVatron~\cite{ROYON},
thanks to the large center-of-mass energy, the inclusive
jet production has been measured up to $p_{t,jet} \approx 500$~GeV.
Moreover, the angular distribution has been studied.
The data 
indicate
better
agreement with predictions using proton structure functions in which
the gluon distribution has been increased at large $x$.

At the near-future programs TeVatron Run II and HERA lumi-upgrade,
more precise tests at highest $p_t^2$ and $Q^2$
will provide a better knowledge of structure functions, which is 
a crucial element for future searches at the LHC.
In some measurements, the dominant errors 
arise due to the lack of accuracy of the theoretical predictions.
State-of-the-art calculations from the theorists' point of view
have been presented~\cite{POETTER}.
A number of NLO programs are available, and theorists are working hard
for the next step, i.e. NNLO programs.
Generally, calculation programs for $\bar p p$ and $e^+e^-$ are
more advanced than $ep$ in terms of parton multiplicity.
Also, only one NLO program for $ep$ can include electroweak couplings.
With high-statistics measurements anticipated, the HERA experimentalists
strongly wish for developments in this field.

\section{Electroweak Physics and high {\boldmath $Q^2$} DIS}
The latest results on electroweak measurements from LEP2~\cite{TERRANOVA}
and TeVatron~\cite{MACHEFERT} have been reviewed.
The $W$-pair production cross-section at LEP2 has been presented
up to a center-of-mass energy of 202~GeV, and the data agree very well
with recent detailed theoretical calculations such as {\tt RacoonWW}.
The latest LEP2-combined $W$-mass measurement gives
$M_W = 80.401 \pm 0.048$~GeV.
The Higgs search has so far yielded no signal, 
leading to
$m_H > 107.7$~GeV.
At TeVatron, a combined measurement of the $W$-mass yields
$M_W = 80.448 \pm 0.062$~GeV.  
The $W$-width is also measured.  
The Run II prospects were presented, including
an expected top-mass measurement with a 3~GeV accuracy
(currently about 5~GeV)
and a $W$-mass precision of 40~MeV per experiment.

\begin{figure}[htb]
 \begin{center}
 \begin{tabular}{p{0.58\textwidth}p{0.37\textwidth}}
     \raisebox{-140pt}{
     \mbox{\epsfxsize=0.6\textwidth
         \epsffile{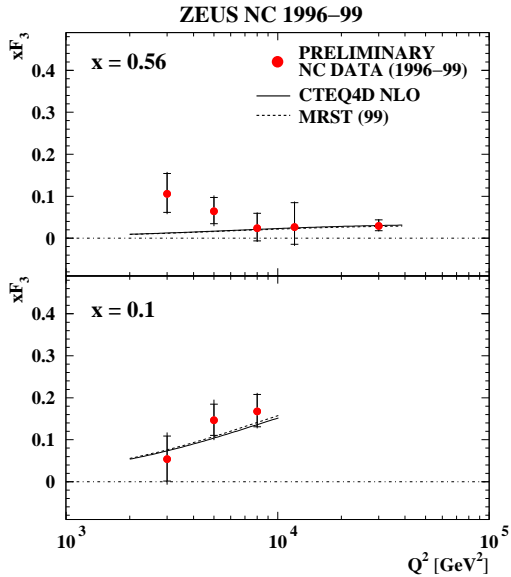}}}
 &
         \caption
         {\small \label{fig:xf3}
            Extracted $xF_3$, plotted as a function of $Q^2$, at
            two different $x$ values.}
 \end{tabular}
 \end{center}
\end{figure}

The latest results on high-$Q^2$ cross-section measurements
in NC and CC DIS processes at HERA were reviewed by H1~\cite{RIZVI}
and ZEUS~\cite{KAPPES}.
Both $e^+p$ and $e^-p$ datasets are available.
In the NC data, the $e^-/e^+$ cross-sections start to separate
at large $Q^2$, where 
$Z$-exchanges 
give a significant contribution.
The difference is essentially $xF_3$, which is proportional to
the difference between the quark and antiquark densities in leading order.
ZEUS has extracted $xF_3$ for $Q^2 > 3000$~GeV$^2$ for the first time
(Fig.~\ref{fig:xf3}).
Although the statistical accuracy is limited, the result is consistent
with the SM prediction with recent parton-density fits.
For the CC process, different electron charge means 
probing different quark flavors.
The $Q^2$-dependence is driven by the $W$-propagator
mass, giving a space-like measurement of $M_W$.
H1 has presented the double differential cross-section $d^2\sigma/dxdQ^2$ from
recent $e^-p$ data, 
as compared to
$e^+p$.  Clearly a higher cross-section
for $e^-p$ than $e^+p$ is seen, and the difference increases as $Q^2$
increases.

Reports
have been made on the future electroweak physics at
the upgraded HERA, from 
both the 
theoretical~\cite{SPIESBERGER}
and the experimental~\cite{LONG} point of view.
Using high-statistics CC and NC data,
constraints of about 50~MeV on $M_W$  in the space-like region
can be obtained; however,
this is true only if the systematic uncertainties can be controlled at 
the 1--2\% level.  
This is a challenge for experimentalists
(e.g. detector energy scale), as well as for theorists
(e.g. reduction of the scale uncertainties in the higher-order calculation).
Moreover,
the error from the parton-density functions has to be substantially
reduced.

One of the important features of 
the HERA upgrade
is the introduction
of longitudinal electron polarization to collider experiments.
It has been shown that high polarization, as well as its accurate
measurement, is essential to many studies, 
e.g. to determine
the vector and axial-vector coupling of light quarks.

\section{{\boldmath $W/Z$} Production, High-{\boldmath $p_t$} leptons}
\label{sec:isolep}

The on-shell prodiction of the $W/Z$ gauge bosons
is an interesting
test of
non-trivial SM calculations which involve gauge cancellations,
resummations and higher-order corrections (both QCD and electroweak).
The recent status of the 
theoretical calculations and the understanding of 
uncertainties has been reviewed~\cite{STIRLING}.
At LEP2, $W$-pair production is predicted to an accuracy of about 0.1\%,
in
very good agreement with the measurements.
For $W$ production at TeVatron, 
the electroweak one-loop corrections
and the NNLO QCD corrections are available.
On the other hand, the calculation for HERA is not as advanced as
those mentioned above, although 
a recent calculation of the QCD correctios
for $ep \to eWX$ process~\cite{NASON} is worth 
noting.
With the upcoming high-statistics productions at the upgraded HERA,
more work on this field is desired.

Experimental results on the QCD aspects of $W/Z$ production at TeVatron
have been presented~\cite{BASSLER}.
The high-statistics $W$ production allows for an accurate comparison of
the $p_t(W)$ distribution with NNLO QCD predictions, and good agreement
is observed.
In addition,
$p_t(Z)$ is compared with the theoretical calculations at both the
low- and high-$p_t$ regions.  At low $p_t$, soft gluon radiation dominates,
requiring resummation of large logarithmic terms and handling of
non-perturbative contributions. The predictions successfully describe
the data over the whole $p_t$ range.

\begin{table}[t]
\caption{Compilation of
events with an isolated high-$p_t$ lepton and missing $p_t$
found by H1 and ZEUS.
The numbers show the observed/expected events for electron ($e$),
muon ($\mu$) and the sum ($e+\mu$).
\label{tab:lepton}}
\vspace{0.2cm}
\begin{center}
\footnotesize
\begin{tabular}{|c|c|c|c|c|c|c|c|}
\hline
\multicolumn{4}{|c|}{H1} &
\multicolumn{4}{|c|}{ZEUS} \\
\hline
{Data period} & $e$ & $\mu$ & $e + \mu $ &
{Data period} & $e$ & $\mu$ & $e + \mu $ \\
\hline
\begin{minipage}{1.8cm}\begin{center}
94-97 $e^+p$\\ (36.5~pb$^{-1}$)
\end{center}\end{minipage}
& 0/0.50 & 5/0.56 & 5/1.06 &
\begin{minipage}{1.8cm}\begin{center}
94-97 $e^+p$\\ (48~pb$^{-1}$)
\end{center}\end{minipage}
& 3/3.5  & 0/2.0  & 3/5.5  \\
\hline
\begin{minipage}{1.8cm}\begin{center}
98-99 $e^-p$\\ (12.8~pb$^{-1}$)
\end{center}\end{minipage}
& 0/0.25 & 0/0.24 & 0/0.49 &
\begin{minipage}{1.8cm}\begin{center}
98-99 $e^-p$\\ (16~pb$^{-1}$)
\end{center}\end{minipage}
& 2/0.8 & 0/0.8 & 2/1.6 \\
\hline
\begin{minipage}{1.8cm}\begin{center}
99-00 $e^+p$\\ (24.9~pb$^{-1}$)
\end{center}\end{minipage}
& 2/0.60 & 1/0.51 & 3/1.11 &
\begin{minipage}{1.8cm}\begin{center}
99    $e^+p$\\ (18~pb$^{-1}$)
\end{center}\end{minipage}
& 2/1.8 & 4/0.9 & 6/2.7 \\
\hline
       $\Sigma$ (74.2~pb$^{-1}$) & 2/1.35 & 6/1.31 & 8/2.66 &
       $\Sigma$ (82~pb$^{-1}$) & 7/6.1 & 4/3.7 & 11/9.8 \\
\hline
\end{tabular}
\end{center}
\end{table}

Closely related to $W$ production, events with an isolated
lepton ($e$ or $\mu$) and missing $p_t$ (attributed to the 
neutrino) have
drawn considerable attention at HERA, mainly due to the muon events
observed by the H1 experiment in the
1994-97 $e^+p$ data.  The rate of these events,
as well as the distribution of the kinematic variables, are largely
different from the SM expectations, where the dominant contribution
comes from the $W$ production.
The ZEUS experiment, on the other hand, observed no such excess and
its results were consistent with the SM.
An update with 1998-99 $e^-p$ data was done by both collaborations
last summer, and no deviation was reported by either experiment.

\begin{figure}[t]
 \begin{center}
   \begin{tabular}{cc}
   \epsfig{figure=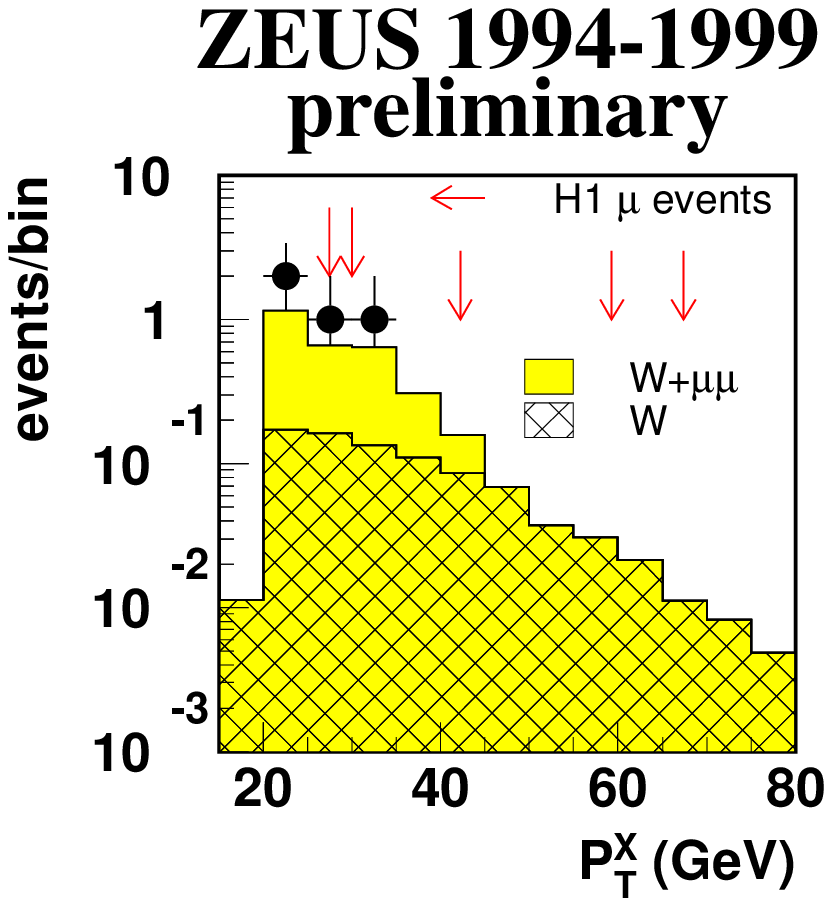,width=0.45\textwidth}
     &
    \epsfig{figure=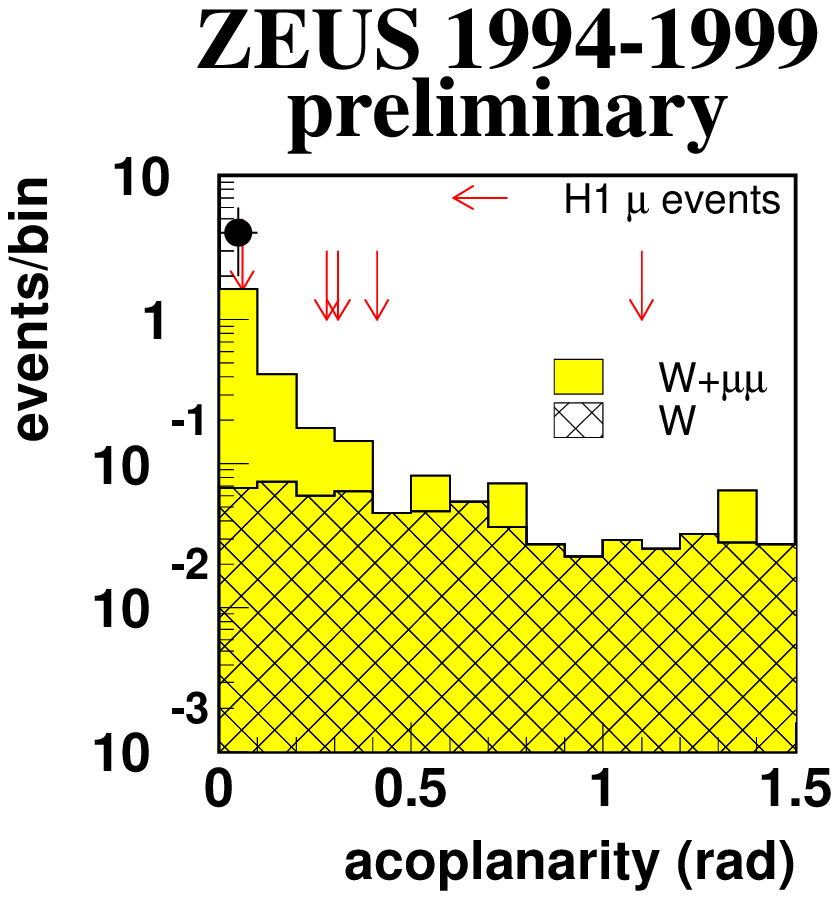,width=0.45\textwidth}
   \end{tabular}
   \caption
         {\small \label{fig:zeusisolep}
            Distributions of the acoplanarity between the muon
            and the hadronic system (left), and of the $p_t$ of
            the hadronic final state $p^X_T$ (right) of the four $\mu$ events
            observed by ZEUS in the 1999 data. 
            The histograms show the SM backgrounds expected in the
            ZEUS analysis for the 1994-99 luminosity.
            The arrows indicate
            the acoplanarity and $p^X_T$ of the five $\mu$ events observed
            by H1 in the 1994-97 data. }
 \end{center}
\end{figure}

In this conference, the results from the 1999-2000 $e^+p$ data
were presented~\cite{MALDEN} from both H1 and ZEUS.
ZEUS finds two electron events and four muon events from the 1999 data, where
1.8 and 0.9 events are expected, respectively.
Although the rate of muon events is higher than expected
in this dataset, 
when combined  with the previous datasets in which
no muon event was observed, the total sample shows good agreement
with the expectations
(7 electron events seen with $6.1\pm0.9$ expected,
and 4 muon events seen with
$3.7\pm0.4$ expected).
Moreover, the events show a back-to-back topology between the muon
and the hadronic system, typical of a SM $\gamma \gamma \to \mu \mu$ process
in which one of the muons has very low $p_t$,
with the $p_t$ of the hadronic final state, $p_T^X$, being not too large.
The distributions of $p_T^X$ and of the $\mu$-hadrons acoplanarity 
of the $\mu$ events are compared to the SM expectations in Fig.~\ref{fig:zeusisolep},
where the $\mu$ events observed by H1 in the 1994-97 $e^+ p$ data are also
indicated as vertical arrows.

From the 1999-2000 data, H1 has observed two electron events and one muon event,
all with fairly large hadronic $p_t$, greater
than 40~GeV.
Summing the electron and muon events from all datasets, the total number
of H1 events in the region of hadronic $p_t > 25$~GeV
is 8, 
with the SM expectation being $2.66\pm0.60$.
The detailed breakdown is shown in Table~\ref{tab:lepton}.
Note that the expected number of events for H1
is generally smaller than for ZEUS, due to
additional cuts to enhance the W events and suppress other SM processes.
Combined efforts towards a detailed understanding of the
differences have started between
the two collaborations.

\section{Parton Densities at high {\boldmath $x$}}

A precise knowledge of the SM expectations is 
a key ingredient for the searches
for new phenomena described in the remaining part of this report.
In general, new physics is already constrained to
appear at the highest accessible scales, where (at
$ep$ and hadronic colliders)  
high-$x$ incoming partons are involved.
Hence,
the precise knowledge of the high-$x$ parton densities in the
proton is crucial for new physics searches.
It was shown~\cite{OLNESS,PAGANIS} 
that the parton densities
in the high-$x$ region are not so well constrained by the existing data.
This is in particular the case for the $d$-quark density,
which is not constrained by the $W$ asymmetries at the
TeVatron~\cite{BASSLER} for  $x > 0.3$.
For higher $x$ values, information on $d(x)$ is obtained
from the $F^d_2$ data, but the size of the nuclear corrections
which have to be applied is unknown. This large uncertainty
on $d(x)$ at high $x$ could be significantly reduced from
the future HERA CC $e^+ p$ data.

\section{Single Top Production}

The cross-section for single top production at the TeVatron II
is $\sim 3.3$ pb. Although this is 3 orders of magnitude smaller
than the $W$+jets production rate, 
it was shown~\cite{BELYAEV} that the extraction of the single top
signal is possible.
This allows to carry out specific measurements,
such as the total top decay width or the
Cabibbo-Kobayashi-Maskawa matrix element $V_{tb}$.
Moreover, single top
production is a gold-plated process to search for new physics.

At LEP and HERA, the SM production cross-section for single top is tiny. However,
higher-order operators 
may allow  a top quark coupling to
a light quark and a gauge boson, which would lead to 
flavor-changing neutral current (FCNC) single top production at LEP2 and HERA.
The cross-sections depend on the size
of the anomalous couplings $\kappa_V$ at the $tqV$ vertices ($V=\gamma, Z$), 
which are constrained
by upper limits of the branching ratio $t \rightarrow V + u/c$.
Single top production followed by $t \rightarrow Wb$ 
has been investigated by the LEP 
experiments~\cite{ALEMANNI}, exploiting the fact that 
the $b$-quark is emitted at a quasi-fixed energy. 
Some 
candidate events
were observed, but no excess over the
SM background (mainly from $WW$ production) was reported; upper limits
on the anomalous couplings were derived,
improving
the TeVatron
bounds on $\kappa_Z$. However, LEP data provide less
sensivity on $\kappa_{\gamma}$, and the full coming LEP2 data should only
marginally improve the bound set by $BR( t \rightarrow \gamma q)$ on
this coupling. 
It is interesting to note that, for $\kappa_{\gamma}$ equal to its
upper bound, the cross-section for single top production at HERA
via $e u \rightarrow e t$ is $\sim $ 1 pb and could thus be observable.
In particular, the leptonic decays of the $W$ would give rise to events
with a high-$p_t$ isolated lepton and a very energetic hadronic
final state, similar to those discussed in section~\ref{sec:isolep}.

\section{Excited Fermions}

\begin{figure}[htb]
 \begin{center}
 \vspace*{-0.7cm}
 \begin{tabular}{p{0.58\textwidth}p{0.37\textwidth}}
     \raisebox{-140pt}{
     \mbox{\epsfxsize=0.6\textwidth
         \epsffile{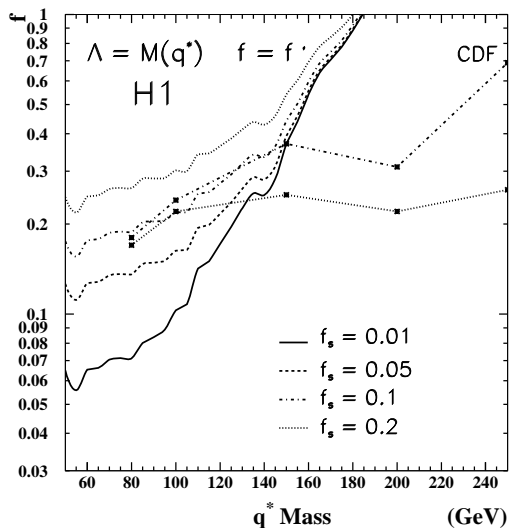}}}
 &
         \caption
         {\small \label{fig:qstar}
              Constraints on $q^*$ assuming $f=f'$ and $\Lambda = M(q^*)$
              for different $f_s$ values, as obtained by the H1 and CDF
              experiments. }
 \end{tabular}
 \vspace*{-0.4cm}
 \end{center}
\end{figure}

Excited fermions ($f^*$) would be a clear evidence for fermion substructure.
At HERA, these could be singly produced via the $t$-channel exchange
of a gauge boson, and would 
subsequently
decay into a SM fermion and a boson $V$.
Collider searches are generally interpreted in the framework of
the phenomenological model~\cite{HAGIWARA}, 
where 
the interactions of $f^*$ with a SM fermion and an electroweak boson
(a gluon) are parameterized via the compositeness scale $\Lambda$
and relative couplings $f$ and $f'$ ($f_s$). \\
Final results from a search for $e^*$, $\nu^*$ and $q^*$ using
the 94-97 $e^+ p$ data have been reported by the H1 Collaboration
at this conference~\cite{COUSINOU}. 
The decays of $f^*$ into $\gamma$, $Z$ and $W$, followed
by the subsequent decay of the boson into $e$, $\mu$, $\nu$ or hadrons
have been investigated, and no deviation from the SM predictions have
been observed. This leads to constraints on the considered model, which
complement the bounds derived by other experiments. In particular,
while the TeVatron sets very stringent bounds~\cite{BERTRAM}
on excited quarks
$q^*$ produced in $qg$ fusion via the coupling constant $f_s$
($ M(q^*) > 760$ GeV for $f=f'=f_s$),
the limits obtained at HERA are better, 
provided
$f_s$ is smaller
than $\sim 0.1$, as illustrated in Fig.~\ref{fig:qstar}.
%


In addition, excited neutrinos $\nu^*$ have been searched for 
by both Collaborations in 
the $e^- p$ data accumulated in 98-99, via their decay 
$\nu^* \rightarrow \nu \gamma$. The preliminary resulting bound improved
significantly the one obtained from the larger statistics $e^+ p$
sample, due to 
the significantly higher
production cross-section. 
Under the
assumption $f / \Lambda = 1 / M(\nu^*)$, $\nu^*$ lighter
than 161 GeV can be excluded at $95 \%$ confidence level.
%

\section{New {\boldmath $eeqq$} interactions}

\subsection{Contact Interactions}

Contact Interactions (CI) can be used to parameterize any new
physics process appearing at an energy scale $\Lambda$ above the
center-of-mass energy $\sqrt{s}$.
At HERA, TeVatron and LEP, $eeqq$ contact interactions would interfere
(constructively or destructively)
with SM processes, 
namely
NC DIS, Drell-Yan and 
$q \bar{q}$ production.

The H1 and ZEUS Collaborations
searched for such distortions in the full $e^+ p$ data collected
between 1994 and 1997~\cite{HERACI}.
An update of these results~\cite{HERACI} was obtained by H1, by including the
$e^- p$ data, which allowed to significantly improve the limits on the
scale $\Lambda$ for some models. The resulting bounds depend on the
CI model and reach 6.4 TeV, 
being competivive
with or better 
than the corresponding limits obtained at the TeVatron from
high-mass Drell-Yan $ee$ events~\cite{BERTRAM}.
In contrast to the more stringent
limits obtained by the LEP experiments, these bounds do not rely
on the flavor symmetry hypothesis.
The CI analysis has been interpreted by the H1 Collaboration
in terms of quark radius, applying a multiplicative form
factor to the SM $d \sigma / d Q^2$. Very heavy new bosons
(leptoquarks) coupling to $e-q$ pairs have also been constrained.
In addition, constraints on models with large extra dimensions where
gravity could become strong at the electroweak scale have been
derived 
for the first time from $eq$ scattering.
The mass scale $M_s$ characterising the effective coupling
of the tower of Kaluza-Klein gravitons with SM fermions 
is constrained to be greater than $\sim 700$ GeV, which is comparable
to limits obtained from the $f \bar{f}$ ($f \neq e$) production at LEP.

\subsection{Leptoquarks}

 Leptoquarks (LQs) are scalar or vector color-triplet bosons
which appear in many extensions of the SM and 
carry both lepton ($L$) and baryon ($B$) numbers.
 At HERA, LQs with
 fermion number $F=3B+L=0$ ($F =2$) could be
 resonantly produced via fusion between the incoming positron
 (electron) and a valence quark coming from the proton. Hence, the $e^+ p$
 and $e^- p$ data are complementary, since they allow to probe 
 different LQ types. LQs can decay back to $e + q$ with a branching
 ratio $\beta$, leading  to NC DIS-like final states.

 \begin{figure}[b]
 \begin{center}
 \vspace*{-0.7cm}
 \begin{tabular}{p{0.58\textwidth}p{0.37\textwidth}}
     \hspace*{-1.5cm}\raisebox{-140pt}{
     \mbox{\epsfxsize=0.75\textwidth
         \epsffile{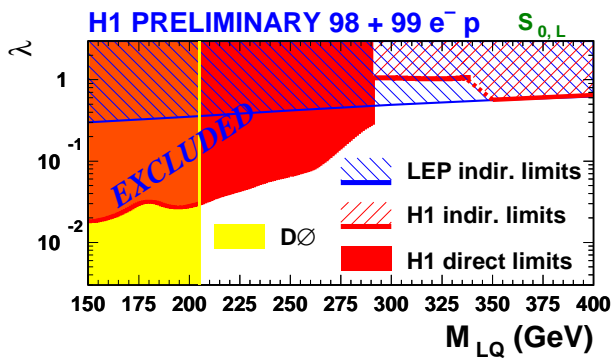}}}
 &
         \caption
         {\small \label{fig:lqlimits}
              Mass-dependent upper bounds on the coupling $\lambda$
              for a $F=2$ LQ decaying
              into $eq$ and $\nu q$ with an equal branching ratio
              of 50 $\%$. Shadowed (hashed) domains are excluded
              by direct (indirect) searches carried out at HERA,
              LEP and TeVatron. }
 \end{tabular}
 \vspace*{-0.1cm}
 \end{center}
\end{figure}

ZEUS and H1 searched for  LQs in the 94-97 $e^+ p$ data
and final results have been published~\cite{HERALQ}.
The H1 analysis makes use of a mass-dependent lower $y_{cut}$
to maximize the significance of a LQ signal, by exploiting
the specific angular distribution of the LQ decay products.
A slight excess of high-$y$
events is observed in the H1 data for invariant masses around 200 GeV,
mainly due to events previously reported in the 94-96 data.
ZEUS performed a general $e-q$ resonance search and also reports
a slight (not statistically compelling)
excess of events at high mass. However, these events are
observed at low-$y$ values, which would not be expected from a
LQ signal. 
%
Both HERA experiments have performed a preliminary analysis of
the $e^- p$ dataset, and no deviation from the SM prediction
has been observed~\cite{FOXMURPHY}.
The $e^+ p$ ($e^- p$) analyses allowed 
stringent bounds to be set on $F=0$ ($F=2$) LQs.
In addition, ZEUS presented a first look at the most recent $e^+ p$ data
accumulated in 1999. No deviation from SM was yet observed in this
small statistics (18 pb$^{-1}$) sample.

The current status on LQs constraints is presented in Fig.~\ref{fig:lqlimits},
for the example case of a $F=2$ LQ decaying to $eq$ and $\nu q$ with
$\beta = 0.5$. The results are shown as mass-dependent upper bounds 
on the coupling $\lambda$ of the LQ to the $eq$ pair. 
At the TeVatron, LQs are mainly pair-produced via the strong coupling, 
which results in $\lambda$-independent mass bounds~\cite{ACOSTA}.
It can be seen in Fig.~\ref{fig:lqlimits} that
the three colliders explore the $(\lambda, M_{LQ})$ plane in a
complementary way and, for an electromagnetic strength of the Yukawa coupling
$\lambda$ (i.e. $\lambda = 0.3$), LQ masses up to 290 GeV can
be ruled out.
Note that for LQs decaying only into $e+q$, a lower mass
bound of 242 GeV can be obtained by combining $D\emptyset$ and CDF
data.

\begin{figure}[htb]
 \begin{center}
 \vspace*{-0.7cm}
 \begin{tabular}{p{0.58\textwidth}p{0.37\textwidth}}
     \raisebox{-140pt}{
     \mbox{\epsfxsize=0.6\textwidth
         \epsffile{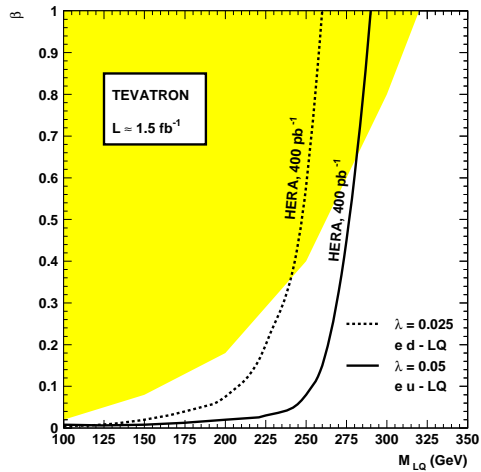}}}
 &
         \caption
         {\small \label{fig:lqfuture}
              Expected mass-dependent sensitivities on the branching ratio
              $\beta$ of a LQ decaying into $eq$ at the
              upgraded HERA and at TeVatron II, indicated as
              full lines and shadowed domain respectively.
              Two typical cases are chosen
              for HERA's sensitivity. }
 \end{tabular}
 \end{center}
\end{figure}

Alternatively, for a fixed value of the coupling $\lambda$, mass-dependent 
constraints on $\beta$ were obtained from the HERA data.
The derived bounds extend beyond the domain excluded by the
TeVatron, for low values of $\beta$~\cite{HERALQ,FOXMURPHY}. The comparison
of the 
expected LQ sensitivity 
at the upgraded HERA and at the TeVatron Run II is
illustrated in Fig. \ref{fig:lqfuture} in the $(\beta, M_{LQ})$ plane, 
which shows the competitivity and complementarity of both facilities.

It is also interesting to note that most recent measurements on
atomic parity violation seem to indicate a $\sim 2 \sigma$ 
discrepancy with the SM predictions, which could be accounted
for by LQ exchange. A global fit of data from HERA, LEP, TeVatron 
and low-energy experiments was performed~\cite{ZARNECKI}, showing
domains in the $(\lambda, M_{LQ}$) plane which could describe
all the datasets.

\section{Lepton Flavor Violation}
\label{sec:lfv}

From the neutrino experiments, it is likely that Lepton Flavor 
Violation (LFV) has been observed in the neutrino sector.
It is thus crucial to look for LFV in the charged lepton
sector as well.
A possible source of LFV at HERA could be the exchange of
LQ bosons 
that couple
both to $e-q$ and
to $\mu -q$ ($\tau -q$) pairs, denoted by
$\lambda_e$ and $\lambda_{\mu}$ ($\lambda_{\tau}$) respectively,
which would lead to $\mu$+jet ($\tau$+jet) final states.
Such final states have been looked for by both H1~\cite{HERALQ}
and ZEUS~\cite{KERGER}, 
finding no events compatible with the
kinematics of LQ-induced $eq \rightarrow lq$ processes.
Upper bounds on the product $\lambda_e \times \sqrt{\beta_{l}}$,
where $\beta_{l}$ denotes the branching
$\beta(LQ \rightarrow l + q)$, have thus been derived as a function
of the LQ mass. 
These searches complement the hunting for $2^{nd}$ and
$3^{rd}$ generation LQs carried out at the TeVatron~\cite{ACOSTA},
for which new results from an improved tagging of
heavy flavors were reported.

\begin{figure}[b]
 \begin{center}
 \vspace*{-0.7cm}
 \begin{tabular}{p{0.55\textwidth}p{0.35\textwidth}}
     \raisebox{-14pt}{
     \hspace*{-3.5cm}\mbox{
     \psfig{figure=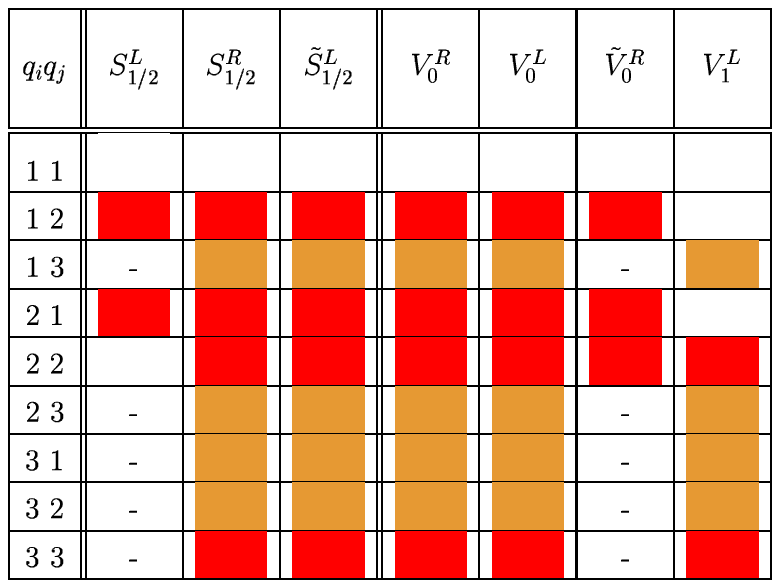,width=0.45\textwidth,angle=-90}
     }}
 &
         \caption
         {\small \label{fig:lfvfuture}
            Future HERA's potential (${\cal{L}} \sim 1$ fb$^{-1}$)
            for LFV LQs coupling to $e-q_i$ and $\tau - q_j$.
            The columns correspond to LQs carrying different
            quantum numbers.
            Dark grey slots indicate cases where HERA will set
            the most stringent bound.
            For light grey slots,
            this should be the case provided that limits from
            rare $B$ decays do not improve by more than a factor
            of two. }  
 \end{tabular}
 \end{center}
 \vspace*{-1cm}
\end{figure}

The case of very high mass ($M_{LQ} \gg \sqrt{s}$)
LFV LQs was also addressed by
H1 and ZEUS. For both $e \leftrightarrow \mu$
and $e \leftrightarrow \tau$ transitions, direct constraints
on such LQs obtained by HERA were compared to the most stringent
indirect bounds.
For some LQ types and  $e q_i \leftrightarrow \tau q_k$
reactions, the HERA limits
extend significantly beyond the reach of other experiments~\cite{KERGER}.
Moreover, the luminosity upgrade will allow HERA to probe most of
these transitions in a domain which is not covered by low-energy
experiments, as shown in Fig.~\ref{fig:lfvfuture}.

On the other hand, $e \leftrightarrow \mu$ transitions are much more
constrained,  by processes such as
$\mu \rightarrow e \gamma$, $\mu \rightarrow 3 e$,
and $\mu-e$
%
%
conversion on heavy nuclei~\cite{KO},
whose rates are experimentally constrained to be
below $ \sim 10^{-11} - 6 \times 10^{-13}$.
Even more stringent bounds are expected in the near future,
which could shed light on the underlying non-SM physics
responsible for LFV.
%
It is important to note that,
in non-supersymmetric models with
massive neutrinos, the amplitudes of these processes
are proportional to the neutrino square mass difference
and thus extremely suppressed.
However, in supersymmetric models they
are only suppressed
by inverse powers of the supersymmetry breaking scale,
and (depending on
the masses and mixings of superparticles) can
lead to rates that are very close to the current
bounds given above. 
Even for universality
at ${M_{GUT}}$, renormalisation effects due to
massive neutrinos will generate off-diagonal sfermion
masses.
What is actually very interesting is the correlation
between the LFV processes in different
supersymmetric theories.
In the MSSM, $\mu \rightarrow e \gamma$,
$\mu \ra 3e$ and $\mu-e$ conversion
all occur via one-loop diagrams, leading to 
approximate relations, e.g.
$\Gamma ( \mu^+ \ra e^+e^+e^-) / \Gamma ( \mu^+ \ra e^+ \gamma)
\approx 6 \times 10^{-3} $.
On the other hand, in the presence of R-parity violation,
$\mu \rightarrow 3 e$ and $\mu-e$ conversion may
occur at tree-level. If, for instance,
only $\lambda_{131}^* \lambda_{231}$
is non-zero, one finds that~\cite{TOBE} 
$\Gamma (\mu \rightarrow e \gamma) / \Gamma (\mu \rightarrow 3e)
= 1\times 10^{-4} ~~{\rm for~~}
m_{\tilde{\nu}_3}=m_{\tilde{e}_R}=100~{\rm GeV},$
a very different prediction from the one
in the MSSM with right-handed neutrinos.

%
%
%
%
%

\section{Supersymmetry searches}

The recent neutrino data seem to strongly indicate
the presence of non-zero neutrino masses, and thus
the existence of {\it beyond the  Standard Model physics}.
Among the various possibilities, supersymmetric theories
are the most promising, since they also provide a solution
to the hierarchy problem and consistent gauge
unification. In this framework,
the simplest model to search for is the so-called
Minimal Supersymmetric Standard Model (MSSM), where
sparticles are only produced in pairs and the lightest
supersymmetric particle (LSP) is stable and neutral,
leading to missing energy signatures.

The limits for selectrons
and smuons obtained at LEP2~\cite{LEPdis} are $m_{\tilde{e}}> 92$ GeV and 
$m_{\tilde{\mu}}> 85$ GeV respectively. For
stau pair production, there exists a small excess
in all four experiments. 
LEP2 also gives bounds for charginos and
neutralinos: $m_{\chi^0} > 37.5 $ GeV and 
$m_{\chi^\pm} > 101$ GeV. Finally, 
searches for squark and gluon pair productions at the TeVatron~\cite{TEVAdis}
give $m_{\tilde{q}} \geq 240$ GeV and
$m_{\tilde{g}} \geq 230$ GeV, while for stop-squarks the
bound drops to 130 GeV due to mixing effects.
These limits will improve during the future TeVatron runs,
while the mass reach will significantly increase
at the LHC \cite{AndyP}.

Supposing that a signal is found,
a fundamental question
is how to deduce the supersymmetric parameters,
in a way as model independent as possible.
This can be done by the combined study of the
total cross-sections and spin correlation 
measurements, and an optimal channel is
chargino pair production in $e^+ e^-$ collisions 
\cite{Jan}.

\begin{figure}[htb]
 \begin{center}
 \begin{tabular}{p{0.58\textwidth}p{0.37\textwidth}}
     \raisebox{-160pt}{
     \mbox{\epsfxsize=0.6\textwidth
         \epsffile{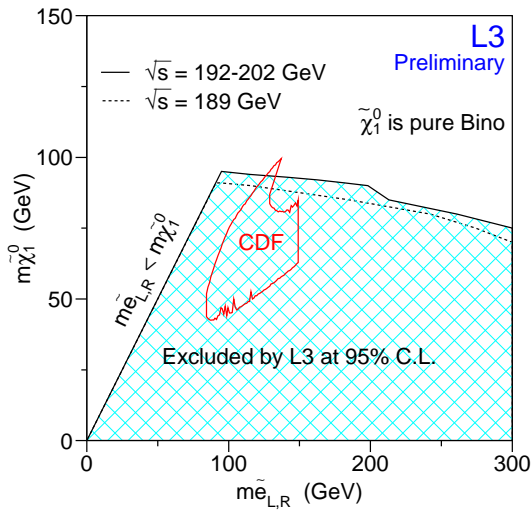}}}
 &
         \caption
         {\small \label{fig:l3gmsb}
              Domain excluded by L3 from searches for pair-produced
              selectrons where 
              $\tilde{e} \rightarrow \chi^0_1 \rightarrow \gamma \tilde{G}$,
              in the plane spanned by the masses of the lightest neutralino 
              $\chi^0_1$ and of the selectron.
              The region of this plane where GMSB models with
              a $\tilde{G}$ LSP could account for the
              $ee \gamma \gamma E_{miss}$ event observed by CDF
              is also represented. }
 \end{tabular}
 \end{center}
  \vspace*{-0.7cm}
\end{figure}

So far, we have been focusing on the minimal supergravity
scenario. However, one can search for alternative theories,
such as those with gauge-med\-ia\-ted~\cite{GauMe} (GMSB) or anomaly-mediated~\cite{RanSun} 
(AMSB) supersymmetry breaking. Unlike supergravity, where symmetry breaking
occurs at $M_{Planck}$ and is communicated via gravitational
interactions to the observable sector, in gauge mediation
this breaking occurs at a scale $\sqrt{F} \approx 100$ TeV and is
communicated via gauge interactions and a so-called
``messenger-sector''. The generic signatures are the existence
of a super-light gravitino $\tilde{G}$ and
processes such as $\chi \rightarrow \gamma \tilde{G}$ or
$\tilde{\ell}^\pm \rightarrow \ell^\pm \tilde{G}$.
So far, no signal has been detected at LEP2 or TeVatron,
leading to the bounds $\sqrt{F} > 217$ GeV and
$m_{\tilde{G}} > 1.1 \times 10^{-5} {\rm eV}$.
Moreover, LEP searches~\cite{GMSBL3} for pair-produced $\tilde{e}$
where $\tilde{e} \rightarrow \chi^0_1 \rightarrow \gamma \tilde{G}$
allow to almost rule out the interpretation of the CDF
$ee \gamma \gamma E_{miss}$ event in GMSB models with
a $\tilde{G}$ LSP, as shown in Fig.~\ref{fig:l3gmsb}.

In AMSB~\cite{RanSun,Ben} theories, supersymmetry breaking is
originated by a rescaling anomaly and the superparticle
spectrum is expressed in terms of beta-functions.
A generic prediction is $M_1 > M_2$, leading to
very light charginos and therefore a characteristic 
associated phenomenology~\cite{FengMor}.

Until now,  we discussed the status of searches
for supersymmetry with the minimal Yukawa sector.
In the most general case, we can also have contributions
violating lepton or baryon number, in combinations such
that the proton is still stable ($R$-parity ($R_p$) violating
supersymmetry). The new operators have the form
$L_iL_j{\bar E}_k, ~L_iQ_j{\bar D_k}$ and
${\bar U_i}{\bar D_j}{\bar D_k}$
where $L$ $(Q)$ are the superfields containing the left-handed leptons (quarks)
doublets,  ${\bar E} ({\bar D}$, ${\bar U}$) those describing
the right-handed singlets, and $i,j,k$ are generation indices.
%
In general, searches for SUSY with $R_p$ violation 
at LEP~\cite{LEPdis} and TeVatron~\cite{TEVAdis}
provide a sensitivity on sparticle
masses which is similar to that obtained with $R_p$
conservation.
%
For HERA, an optimal search channel for $R_p$-violating
supersymmetry is through resonant single squark production
via an $L_1Q\bar{D}$ operator \cite{FOXMURPHY}.
This process is similar to single scalar leptoquark
production, however in $R_p$ violation we also have to consider
the cascade decays
of squarks to neutralinos and charginos.
Especially for
$L_1Q_{2,3}\bar{D}_1$, one finds that the HERA limits can
be better than those from alternative processes, 
such as atomic parity violation, for squark masses
below 210 GeV. 
%
Finally, in the case that two $R_p$-violating operators involving
different lepton flavours are
large, HERA gives strong bounds on products
of $(L_1Q\bar{D}) (L_3Q\bar{D})$ operators,
in analogy with the results for scalar
leptoquarks \cite{KERGER} reported in section~\ref{sec:lfv}.

\section{Higgs searches}

In the SM a Higgs particle, $H$, 
is predicted by the spontaneous breaking of the
electroweak symmetry, while the minimal supersymmetric extension of
the theory requires two Higgs doublets, leading
to two neutral CP-even states
($h,H$), one neutral CP-odd ($A$) and two charged ones ($H^\pm$). 
At tree-level, the mass of the lightest scalar Higgs boson $h$ 
is restricted to be
$ m_h < M_Z$, however, radiative corrections 
drive this bound to $m_h \leq 135$ GeV.
Thus, it is a generic prediction that supersymmetry leads
to a light Higgs.

The direct SM Higgs search at LEP2 leads
to a lower bound~\cite{TERRANOVA} of $108$ GeV.
At the upgraded TeVatron experiment,
the SM Higgs can be mainly produced via gluon fusion $gg\to H$ and the
Drell--Yan like production $q\bar q \to W^* \to WH$.
However, due to QCD backgrounds, 
the best channel is the Drell--Yan like process
\cite{PRatoff,MSpira}.
Higgs discovery may also be
possible via the $H\to \tau^+\tau^-$ decay in $H+W/Z$,
while  $gg\to H\to W^*W^* \to
\ell \nu jj, \ell^+\ell^- \nu \bar\nu$ is also distinguishable
owing to strong angular
correlations among the final state leptons.
A discovery of the SM Higgs boson at the upgraded TeVatron
with ${\cal L} \approx 30 {\rm ~fb}^{-1}$
might be possible for $m_H \leq 125$ GeV and for
$155 \leq m_H \leq 175$ GeV, 
while for ${\cal L} \approx 10 {\rm ~fb}^{-1}$
a $95\%$ confidence level limit for $m_{H} < 180$ GeV
can  be provided.
Finally, 
at the LHC, an important channel (not good for TeVatron)
will be  $gg\to H\to ZZ^*\to 4\ell^\pm$.

 \begin{figure}[htb]
 \begin{center}
 \vspace*{-0.7cm}
 \begin{tabular}{p{0.58\textwidth}p{0.37\textwidth}}
     \raisebox{-160pt}{
     \mbox{\epsfxsize=0.6\textwidth
         \epsffile{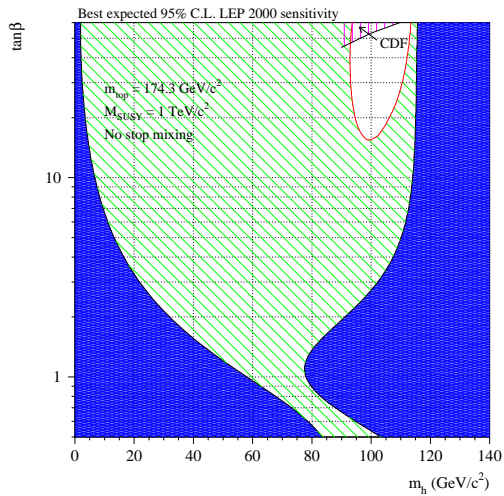}}}
 &
       \caption
         {\small \label{fig:susyhiggses}
              Projected LEP2 contours of the 95 $\%$ CL exclusion
              limits for MSSM Higgs sector parameters, as a function
              of $\tan \beta$ and $m_h$.
              The contours shown assume no stop mixing, leading
              to the maximal expected LEP coverage.
              The dark shaded region is theoretically forbidden.
              The domain 
              currently  excluded by CDF is also represented. }
 \end{tabular}
 \end{center}
\end{figure}

In the case of supersymmetry, 
the light scalar Higgs boson $h$ can be
mainly produced via $q\bar q\to
W/Z + h$ analogously to the SM case.
However, for large $\tan\beta$, the
channels $q\bar q, gg\to b\bar b + h/A$ become
important due to the large $b$-Yukawa couplings
\cite{MSpira}.
%
A search for $b \bar{b} \Phi$ ($\Phi = h,H,A$)
followed by $\Phi \rightarrow b \bar{b}$ has been
performed by CDF and the results were converted into
constraints in the plane $(\tan \beta, m_h)$ which
already cover a region which is not accessible at LEP,
as shown~\cite{JANOT} in Fig.~\ref{fig:susyhiggses}.
In Run II the TeVatron will beautifully complement 
LEP by filling in a much larger part this region
not covered by LEP.

Charged Higgs boson can also be searched for
\cite{PRatoff,MSpira},
while the effects of radiative corrections turn out to be
significant and up to $100\%$ in certain
cases~\cite{MSpira}.

Finally, there also exists the 
possibility that no light scalar Higgs field exists,
and symmetry breaking occurs due to the non-zero vacuum 
expectation values
of composite operators, in the presence of
strongly interacting gauge bosons at higher energies
\cite{Jeff}.
So far, the only probe for such theories comes from
electroweak precision data. However one may hope to
probe at the LHC or the linear collider 
anomalous four-boson effective vertices, which will be a strong
indication for this type of physics.

 \section{Conclusions}
%
In this workshop, the status of physics at highest $Q^2$ and $p_t^2$
was reviewed. Up to this extreme phase space, where all experiments
face the limit of statistics, the SM predictions seem to 
beautifully describe
the cross-sections over 
several orders of magnitude.
These tests
rather provide valuable information which constrains
the parton-density functions of the proton and the photon.
In the near-future programs
of the HERA upgrade and the TeVatron Run II, substantial increase of statistics
will help to further extend the 
available 
phase space, while in some measurements
the developments in the precise theoretical calculation are becoming
a crucial path for reducing errors.

Searches for hints of new physics are extensively carried out at 
HERA, LEP2 and TeVatron,
and so far no definite signal has been observed.  The complementarity
of the colliders has been demonstrated in various searches, and some
new possibilities were discussed.  There are several $2 \sigma$-level
``anomalies" seen,
to be confirmed or disconfirmed with more data to come.

The DIS physics (which includes in a wider sense both lepton-parton and
parton-parton scattering at short distance) will still play
a major role in the next decade of particle physics. 
The ongoing physics study on how to search for 
Higgs and SUSY particles at the LHC is closely 
related to the outcome of the HERA and TeVatron
runs in next few years, 
and these will definitely be an important topic at the ``DIS 2010" workshop.



\end{document}